\documentclass[11pt,a4paper]{article}
\pdfoutput=1

\usepackage{jheppub}
\usepackage[rgb]{xcolor}
\usepackage{color}
\usepackage{amsmath,amsfonts,mathrsfs,graphicx,bbm,dsfont,booktabs,mathtools,braket,lmodern,tikz}

\DeclareMathAlphabet{\mathpzc}{OT1}{pzc}{m}{it}

\def\L{{\mathbb L}}

\newcommand{\bea}{\begin{eqnarray}}
\newcommand{\eea}{\end{eqnarray}}
\def\be{\begin{equation}}
\def\ee{\end{equation}}
\newcommand{\bei}{\begin{itemize}}
\newcommand{\eei}{\end{itemize}}
\newcommand{\bee}{\begin{enumerate}}
\newcommand{\eee}{\end{enumerate}}

\def\a {\alpha}
\def\b {\beta}

\def\p{\phi}

\def\eps{\epsilon}

\def\ov{\over}

\def\pa {\partial}

\def\L{\mathscr L}

\newcommand{\alg}[1]{\mathfrak{#1}}

\newcommand{\psu}{\alg{psu}}

\def\ads{{\rm AdS}_5\times {\rm S}^5}

\def\ads{{\rm AdS}_5\times {\rm S}^5}

\def\am{{\rm am}}
\def\am0{{\rm am}_0}

\expandafter\def\expandafter\bfseries\expandafter{\bfseries\ifmmode\else\boldmath\fi}
\expandafter\def\expandafter\mdseries\expandafter{\mdseries\ifmmode\else\unboldmath\fi}
\expandafter\def\expandafter\normalfont\expandafter{\normalfont\ifmmode\else\unboldmath\fi}

\definecolor{grey}{rgb}{0.4,0.4,0.5}
\definecolor{darkgreen}{rgb}{0,0.5,0}
\definecolor{darkred}{rgb}{0.6,0.0,0}
\definecolor{lightbrown}{rgb}{1,0.9,0.8}
\definecolor{brown}{rgb}{0.6,0.3,0.3}
\definecolor{darkblue}{rgb}{0,0,0.8}
\definecolor{darkmagenta}{rgb}{0.5,0,0.5}

\title{Deformed Neumann model from\\
spinning strings on $\left(\ads\right)_\eta$}

\author[a,1]{Gleb Arutyunov}
\note{Correspondent fellow at Steklov Mathematical Institute, Moscow.}
\author[a]{and Daniel Medina-Rincon}

\affiliation[a]{Institute for Theoretical Physics and Spinoza Institute, Utrecht University, Leuvenlaan 4, 3584 CE Utrecht, The Netherlands}

\emailAdd{g.e.arutyunov@uu.nl}
\emailAdd{d.r.medinarincon@students.uu.nl}

\abstract{We show that bosonic spinning strings on the $\eta$-deformed $\ads$ background 
are naturally described as periodic solutions of a novel finite-dimensional integrable system which can be viewed as a deformation of the celebrated Neumann model. 
For this deformed model we find the Lax representation
and the analogue of the Uhlenbeck integrals.}

\begin{document}

\begin{flushright}\small{ITP-UU-14/17\\SPIN-14/19}\end{flushright}

\maketitle

\section{Introduction}
Spinning string solutions have played a prominent role in recent developments concerning the integrable structure of the planar AdS/CFT system \cite{M}, see for instance \cite{Frolov:2002av}-\cite{Arutyunov:2009ga}. They would not only allow one to test the integrability of the string sigma-model, to match the higher conserved charges of gauge and string theory \cite{Arutyunov:2003rg}, but also to predict a universal integrability structure such as the Quantum String Bethe Ansatz \cite{Kazakov:2004qf,Arutyunov:2004vx}. As is known \cite{Arutyunov:2003uj,Arutyunov:2003za}, in the simplest setting bosonic rigid spinning strings in the $\ads$ space-time are naturally described as periodic solutions of the finite dimensional integrable system due to C. Neumann  \cite{Neumann}. Historically, this system was one of the first known integrable models and, in fact, it was the first significant problem of mechanics to be solved by hyperelliptic functions.  In spite of the fact that the string sigma-model on $\ads$ is two-dimensional with $\tau$ being the world-sheet time and $\sigma$ a (periodic) spatial coordinate, specifying the rotating string ansatz leads to a complete decoupling of $\tau$ from the equations of motion, so that one obtains a finite-dimensional mechanical model  with $\sigma$ playing the role of time. Physically, the Neumann model describes an $N$-dimensional harmonic oscillator with its motion restricted on an $(N-1)$-dimensional sphere. Although integrability of the model was already known to Neumann and Jacobi, the integrals of motion were discovered almost a hundred years later  by K. Uhlenbeck in her work on the description of harmonic maps into spheres \cite{Uhlenbeck}. This finding allowed the model to be put in the framework of the Liouville theorem and to solve it, for instance, by the method of separation of variables, {\it cf.} \cite{BabelonBernardTalon}.  

\smallskip

As was recently shown, the string sigma-model on $\ads$ admits a deformation which preserves its integrability \cite{Delduc:2013qra}\footnote{Some earlier and related work in the context of  AdS/CFT on sigma-model deformations  is \cite{Cherednik:1981df}-\cite{Crichigno:2014ipa}. }. This deformation is governed by a real parameter $\eta$ and, therefore, we will refer to the corresponding background as $(\ads)_{\eta}$, for which the $B$-field and the metric were recently found in \cite{Arutyunov:2013ega}\footnote{It remains unknown if the $(\ads)_{\eta}$-metric and the $B$-field can be lifted to the full solution of IIB supergravity.}. As was already mentioned, this theory (sigma-model) is classically integrable and the perturbative two-body S-matrix has been computed in \cite{Arutyunov:2013ega}. This S-matrix appears to coincide with the large tension limit of the exact S-matrix which respects the $q$-deformed centrally extended quantum supersymmetry algebra $\psu_q(2|2)\oplus \psu_q(2|2)$. Among other recent developments extending and generalizing earlier work on the string sigma-model deformations with $q$ being a root of unity \cite{Hoare:2011wr} -\cite{vanTongeren:2013gva}, we mention the construction of the corresponding thermodynamic Bethe equations for the accompanying mirror model, which should encode the spectrum of the sigma-model on 
$(\ads)_{\eta}$ as well as the new mirror duality phenomenon  \cite{Arutynov:2014ota,Arutyunov:2014cra}.

\smallskip

In the present paper we put forward an interesting integrable deformation of the Neumann model which emerges naturally from strings spinning 
in the $\eta$-deformed $\ads$ space-time.  Since the $\eta$-deformed metric has six ${\rm U}(1)$ isometries, just as in the case of the usual $\ads$, it is possible to impose 
the spinning string ansatz, where the string embedding coordinates corresponding to the isometry directions are chosen to be aligned with the world-sheet time $\tau$. 
As was explained in \cite{Delduc:2013qra}, the sigma-model on the $\eta$-deformed background is integrable and we will show here that the corresponding $8\times8$-matrix Lax representation admits a reduction on the spinning string ansatz, thereby rendering the corresponding model integrable in the Liouville sense. In order not to overload our considerations with unnecessary details, we restrict ourselves to strings spinning in the deformed five-sphere while being at the center of the deformed ${\rm AdS}$ background. A generalization to spinning motion in the AdS-like part of the background metric is straightforward and will be explained elsewhere. As motion is restricted to the deformed sphere, the Lax representation for the corresponding deformed Neumann model comes naturally in terms of $4\times 4$-matrices.

\smallskip

In principle, having a Lax representation for a finite-dimensional integrable system it is straightforward to exhibit integrals of motion as they are 
simply given by ${\rm Tr}(L^k)$, where $L$ is the Lax matrix.   Integrals of motion form, however, a ring, and therefore the integrals emerging as traces of powers of the Lax matrix might not be ``elementary" in the sense that they might be built from much simpler conserved blocks. Indeed, already the usual Neumann model
admits Lax representations in terms of either $2\times 2$ or $3\times 3$, or even $4\times 4$-matrices, where in the latter case the integrals  ${\rm Tr}(L^k)$
are not at all elementary, being intricate algebraic combinations of elementary Uhlenbeck integrals. The main effort of the present work is to use the $4\times 4$-matrix Lax representation inherited from the spinning string ansatz to disclose the elementary conserved quantities -- the deformations of the Uhlenbeck integrals -- arising from strings spinning in the $\eta$-deformed background. For the reader who is interested in the main results, we point out the formulae 
(\ref{Ffinal1})-(\ref{Ffinal3}) which give the conserved quantities for the $\eta$-deformed Neumann model. The deformation parameter $\varkappa$ used there is related to $\eta$ by:
$$ \varkappa= \frac{2\eta }{1-\eta^2}\, .$$ 
It is important to note that our expressions $\mathscr{F}_{i}$ extend the Uhlenbeck integrals for the usual Neumann model as an expansion in the deformation parameter $\varkappa$ up to (and including) terms of order $\varkappa^4$; while the Hamiltonian, being a special combination of the $\mathscr{F}_{i}$, is expanded up to the order $\varkappa^2$ only.

\smallskip

The paper is organized as follows. In section \ref{section2} we recall the basic  points about the Neumann model including its Lagrangian formulation and its Hamiltonian formalism in terms of Dirac brackets. Section \ref{section3} is devoted to the spinning strings ansatz in the $(\ads)_{\eta}$-background. In section \ref{section4} we discuss the $4\times 4$-matrix Lax representation for spinning strings in the $\eta$-deformed background, and in section \ref{section5} we construct the associated integrals of motion in unconstrained coordinates $(r,\xi)$. In section \ref{section6} the cousins of the Uhlenbeck integrals for the deformed model are presented in constrained coordinates $x_{i}$ and their involutive property is verified, constructing in this way the Dirac bracket Hamiltonian formalism for the deformed Neumann model. Finally, in the conclusions a few interesting problems for further study are outlined.

\section{The Neumann Model}
\label{section2}
\subsection{Lagrangian Formulation}
As was mentioned earlier, the Neumann model is a well known integrable system, representing a harmonic oscillator constrained to move on a $(N-1)$-dimensional unit sphere \cite{BabelonBernardTalon}. The Lagrangian for this system is given by:

\begin{equation}
L = \frac{1}{2}\sum\limits_{i = 1}^N {\left( {\dot x_i^2 - \omega _i^2x_i^2} \right)}  + \frac{\Lambda }{2}\left( {\sum\limits_{i = 1}^N {x_i^2 - 1} } \right),
\label{eq1}
\end{equation}
where $\Lambda$ plays the role of a Lagrangian multiplier, while $x_{i}$ and $\omega_{i}$ correspond to the coordinates and angular frequency along the direction $i$. The Euler-Lagrange equations are given by:
\begin{equation}
{{\ddot x}_i} =  - \omega _i^2{x_i} + \Lambda {x_i}\ .
\label{eq2}
\end{equation}
For this system the Lagrangian multiplier can easily be obtained from the equations of motion and the constraint $\sum\nolimits_i {x_i^2} =1$, yielding as a result:
$$\Lambda  = \sum\limits_{i = 1}^N {\left( {\omega _i^2x_i^2 - \dot x_i^2} \right)} .$$
Thus, the dynamics are given by the following non-linear equations of motion:
\begin{equation}
{{\ddot x}_i} =  - \omega _i^2{x_i} + {x_i}\sum\limits_{j = 1}^N {\left( {\omega _j^2x_j^2 - \dot x_j^2} \right)} .
\label{eq3}
\end{equation}

\subsection{Hamiltonian Dirac Bracket Formulation}
\label{DiracFormulation}
The Neumann model is usually studied in the Hamiltonian formalism and therefore we will introduce it briefly. In this formalism the canonical momentum and the (unconstrained) Hamiltonian are given by:
$${\pi_i} = \frac{{\partial L}}{{\partial {{\dot x}_i}}} = {\dot x_i},$$
\begin{equation}
H = \frac{1}{2}\sum\limits_{i = 1}^N {\left( {\pi_i^2 + \omega _i^2x_i^2} \right)}.
\label{eq4}
\end{equation}
The constraint in the $2N$-dimensional phase space is expressed as:
\begin{equation}
\sum\limits_{i = 1}^N {x_i^2}=1,\ \ \ \ \ \ \  \sum\limits_{i = 1}^N {{x_i}{\pi_i}}=0\ . 
\label{constrainsundeformed}
\end{equation}
Due to the second constraint, it is necessary to use the Dirac bracket formalism obtained from the canonical structure $\{\pi_{i},x_{j}\}=\delta_{ij}$. This corresponds to:
\begin{equation}
\{\pi_{i},\pi_{j}\}_{D}=x_{i}\pi_{j}-x_{j}\pi_{i},\ \ \ \ \ \{\pi_{i} ,x_{j} \}_{D}=\delta_{ij}-x_{i}x_{j},\ \ \ \ \ \{x_{i},x_{j}\}_{D}=0.
\label{Diracbrackets}
\end{equation}
The solution to the $N$-dimensional Neumann model and its (Liouville) integrability is due to the existence of $N$ integrals of motion $F_{i}$ first found by Uhlenbeck in \cite{Uhlenbeck}, which satisfy:

\begin{equation}
F_{i}=x_i^{2}+\sum_{j\neq i}\frac{J_{ij}^{2}}{\omega_i^2-\omega_j^2}\ ,\ \ \ \ \ \ \ \sum_{i=1}^{N}F_{i}=1,\ \ \ \ \ \ \ \{F_{i},F_{j}\}_{D}=0,
\label{undeformedFproperties}
\end{equation}
where $J_{ij}=x_{i}\pi_{j}-x_{j}\pi_{i}$. From the equation in the middle of \eqref{undeformedFproperties}, we see that only $N-1$ Uhlenbeck integrals are independent. In particular, we will be interested in the case $N=3$, thus, out of the 3 integrals $F_{i}$ only 2 will be independent. For a general $N$, the Hamiltonian of the Neumann model can also be written as a linear combination of the Uhlenbeck integrals:
\begin{equation}
H=\frac{1}{2}\sum_{i=1}^{N}\omega_i^2 F_{i}.
\label{undeformedhamiltonianF}
\end{equation}
By explicit substitution of the Uhlenbeck integrals $F_{i}$ in the above equation, the Hamiltonian in the Dirac bracket formalism can be written as:
\begin{equation}
H=\frac{1}{4}\sum_{i\neq j}J_{ij}^{2} +\frac{1}{2}\sum_{i}\omega_i^2 x_{i}^{2}\ ,
\label{alternativeH}
\end{equation}
which coincides with the one in equation \eqref{eq4} by using the constraints of equations \eqref{constrainsundeformed}.

\section{Spinning strings on $\eta$-deformed $\ads$}
\label{section3}
The Lagrangian for the sigma-model describing bosonic strings in the $\eta$-deformed background was found in \cite{Arutyunov:2013ega}. Without loss of generality in this work we will restrict our attention to its part corresponding to the deformed five-sphere\footnote{Concerning the fulfillment of the Virasoro constraints, just as in the undeformed case, on the spinning string ansatz the only non-trivial Virasoro constraint will subsequently relate the conserved energies of two Neumann models corresponding to the deformed sphere and deformed AdS space, respectively.}:   

\bea
\L &=&{1\ov2}\, \eta^{\a\b}\Bigg(\frac{\pa_\a \p\pa_\b \p
  \left(1-r^2\right)}{1+\varkappa ^2 r^2}+\frac{\pa_\a r\pa_\b r
  }{ \left(1-r^2\right) \left(1+\varkappa ^2 r^2\right)}
  +\frac{\pa_\a \xi\pa_\b \xi \,  r^2}{1+ \varkappa ^2 r^4 \sin ^2\xi}\label{L}  \\
  \nonumber
   &&\qquad+\frac{\pa_\a \p_1\pa_\b \p_1  r^2 \cos ^2\xi }{1+ \varkappa ^2
   r^4 \sin ^2\xi } +\pa_\a \p_2\pa_\b \p_2\,   r^2 \sin^2\xi \Bigg)+{1\ov2} \varkappa \, \eps^{\a\b}\frac{ r^4 \sin 2 \xi }{1+ \varkappa ^2 r^4 \sin^2\xi}\pa_\a\p_1\pa_\b\xi\\
\nonumber
 &&+\varkappa\epsilon^{\alpha\beta}\frac{ \partial_{\alpha}r\partial_{\beta}\phi\ r}{1+\varkappa^{2}r^{2}}.
\eea
Here the world-sheet metric $\eta^{\a\b}$ is chosen to be Minkowski  and the term with $\eps^{\a\b}$ represents a contribution of the $B$-field. In comparison to the original formulation \cite{Arutyunov:2013ega} we changed the overall scale of $\L$ and included an additional term at the end, this term is a total derivative and therefore does not affect the dynamics of the system. It is worth nothing that this total derivative appears naturally in the derivation of the $\eta$-deformed Lagrangian presented in \cite{Arutyunov:2013ega} and it is included here to make calculations simpler.

Obviously, the Lagrangian (\ref{L}) exhibits three isometries corresponding to shifts of the angles $\phi$, $\phi_1$ and $\phi_2$. This allows one to consider the following ansatz 
for a solution describing strings spinning in three different directions with angular velocities $\omega_1$, $\omega_2$ and $\omega_3$:
\bea
\p_1=\omega_1\tau\, , ~~~\p_2=\omega_2\tau\, , ~~~\p=\omega_3\tau\, ,~~~r\equiv r(\sigma)\, , ~~~\xi\equiv \xi(\sigma)\, ,
\label{spininganzats}
\eea
where $\tau$ and $\sigma$ are the world-sheet time and spatial coordinates, respectively. Since we are dealing with closed strings we assume that 
$r$ and $\xi$ are periodic functions of $\sigma$ with period $2\pi$. Substituting this ansatz into (\ref{L}), we obtain:
\bea\nonumber
\L&=&{1\ov2}\, \Bigg[
\frac{r'^2  }{ \left(1-r^2\right) \left(1+\varkappa ^2 r^2\right)}+\frac{r^2\xi'^2+\varkappa \omega_1 r^4\xi'\sin 2\xi }{1+ \varkappa ^2r^4 \sin ^2\xi }\\
&&\qquad\qquad\qquad
- \frac{\omega_1^2 r^2\cos^2\xi}{1+ \varkappa ^2r^4 \sin ^2\xi}-\omega_2^2r^2\sin^2\xi-\frac{\omega_3^2
  \left(1-r^2\right)}{1+\varkappa ^2 r^2}-\frac{2\varkappa \omega_{3}r\ r'}{1+\varkappa^{2}r^{2}}\Bigg]\, \label{Lagrangianrxi} .\eea
 This is a Lagrangian for a mechanical system where $\sigma$ plays the role of a time variable; prime denotes a derivative with respect to $\sigma$.\footnote{Another mechanical model arises in the particle limit, where all variables are assumed to be $\sigma$-independent. This model describes geodesic motion on the corresponding $\eta$-deformed background and it represents an integrable 
 deformation of the Rosochatius system, as shown in appendix A.} To make further progress, it is convenient to make a change of variables: 
 $$
 r=\sqrt{x_1^2+x_2^2}\, , ~~~~~\xi={\arctan}\frac{x_2}{x_1}\, ,
 $$
 upon which the Lagrangian acquires the form:
\begin{align}
 \label{lagx}
\L=&\frac{1}{2}\bigg[ - \frac{{\omega _1^2x_1^2}}{{1 + {\varkappa ^2}x_2^2\left( {x_1^2 + x_2^2} \right)}} - \frac{{\omega _3^2x_3^2}}{{1 + {\varkappa ^2}\left( {x_1^2 + x_2^2} \right)}} - \omega _2^2x_2^2+ \frac{{2\varkappa {\omega _1}{x_1}{x_2}\left( {{x_1}{{\dot x}_2} - {x_2}{{\dot x}_1}} \right)}}{{1 + {\varkappa ^2}x_2^2\left( {x_1^2 + x_2^2} \right)}}\\
&+ \frac{{{{\left( {{x_2}{{\dot x}_1} - {x_1}{{\dot x}_2}} \right)}^2}}}{{\left( {x_1^2 + x_2^2} \right)\left( {1 + {\varkappa ^2}x_2^2\left( {x_1^2 + x_2^2} \right)} \right)}} + \frac{{2\varkappa {\omega _3}{x_3}{{\dot x}_3}}}{{1 + {\varkappa ^2}\left( {x_1^2 + x_2^2} \right)}} + \frac{{{\dot x}_3^2}}{{\left( {x_1^2 + x_2^2} \right)\left( {1 + {\varkappa ^2}\left( {x_1^2 + x_2^2} \right)} \right)}} \bigg],\nonumber
\end{align}
here we introduced the variable $x_3$ which is not independent but rather subject to the constraint: 
$$ x_1^2+x_2^2+x_3^2=1.$$ 
Using this constraint and its derivative, one can check that in the limit of $\varkappa\rightarrow0$ the Lagrangian of \eqref{lagx} (plus the constraint) reduces to that of the $N=3$ Neumann model, namely, the Lagrangian from equation \eqref{eq1}. In this way the results of \cite{Arutyunov:2003uj}, which where obtained for the undeformed $\ads$ background, are recovered.

\section{Lax Representation from $\eta$-deformed $\ads$}
\label{section4}
The Lax connection for $\eta$-deformed $\ads$ was introduced in \cite{Delduc:2013qra}, this was done by means of vectors $J_{\alpha }$, $\widetilde{J}_{\alpha }$, and their projections $J_{-}^{\alpha}=P^{\alpha\beta}_{-}J_{\beta}$ and $\widetilde{J}_{+}^{\alpha}=P^{\alpha\beta}_{+}\widetilde{J}_{\beta}$; where $P_{\pm}^{\alpha\beta}=\frac{1}{2}(\eta^{\alpha\beta}\pm\epsilon^{\alpha\beta})$. In our case, since we are only interested in the bosonic part, these vectors are given by:
\begin{equation}
J_{\alpha } = -\frac{1}{1 - \varkappa R_{\mathfrak{g}} \circ P_{2}}\left(A_{\alpha} \right),\ \ \ \ \ \ \ \ \ \ \ \ \ \ \widetilde{J}_{\alpha } = -\frac{1}{1 + \varkappa R_{\mathfrak{g}} \circ P_{2}}\left(A_{\alpha} \right).
\label{definitionJs}
\end{equation}
Here we have used $A_{\alpha}=-\mathfrak{g}^{-1}\partial_{\alpha}\mathfrak{g}$, being $\mathfrak{g}=\mathfrak{g}(\tau,\sigma)$ a bosonic coset representative of ${\rm SU}(2,2)\times {\rm SU}(4)/{\rm SO}(4,1) \times {\rm SO}(5)$. The expressions for $J_{\alpha } $ and $\widetilde{J}_{\alpha }$ used here differ from those used in \cite{Delduc:2013qra} by a minus sign, this is due to our definition of the currents $A_{\alpha}$. The action of the operator $R_{\mathfrak{g}}$ on $M\in \mathfrak{psu}(2,2|4)$, as described in \cite{Delduc:2013qra,Arutyunov:2013ega}, is given by:
$$R_{\mathfrak{g}}=\mathfrak{g}^{-1}R(\mathfrak{g}M\mathfrak{g}^{-1})\mathfrak{g} ,$$
being $R$ a linear operator on $\mathfrak{psu}(2,2|4)$ that satisfies the modified classical Yang-Baxter equation, and whose action on an arbitrary $8\times 8$ matrix is given by:

$$R(M)_{ij}=-i\epsilon_{ij} M_{ij}, \ \ \ \ \  \epsilon_{ij}= \left\{
	\begin{array}{ll}
		\ \ 1\ \ \text{if}\  i<j \\
		\ \ 0\ \ \text{if} \ i=j\\
 		-1\ \ \text{if} \ i>j
	\end{array}
\right. .$$
In order to calculate $J_{\alpha }$ and $\widetilde{J}_{\alpha }$ for the case of the spinning strings ansatz of equation \eqref{spininganzats}, it is necessary to invert the operator $1- \varkappa R_{\mathfrak{g}} \circ P_{2}$ (for $\widetilde{J}_{\alpha }$ one has to invert $1+ \varkappa R_{\mathfrak{g}} \circ P_{2}$, which is equivalent to doing $\varkappa\rightarrow-\varkappa$ on the result for $J_{\alpha}$) and to choose a coset representative $\mathfrak{g}$. For this we use the same $\mathfrak{g}$ as used in \cite{Arutyunov:2013ega}, where it is also shown how to invert the operator $1 - \varkappa R_{\mathfrak{g}} \circ P_{2}$, and then we substitute the ansatz of equation \eqref{spininganzats}.
For the bosonic case the ${\rm AdS}$ and sphere parts of every $8\times 8$ matrix can be treated separately, thus, from now on when referring to matrices $J_{\alpha}$ and $\widetilde{J_{\alpha}}$ we will refer only to their $4\times 4$ part associated with the sphere since this is the part relevant for the reduction to the Neumann model.

The Lax connection for $\eta$-deformed $\ads$ proposed in \cite{Delduc:2013qra} is constructed by means of vectors $L^{\alpha}_{+}$ and $M^{\alpha}_{-}$, for the bosonic case these vectors can be written as:
\begin{align}
L^{\alpha}_{+}&=\widetilde{J}_{+}^{\alpha (0)}+\lambda^{-1}\sqrt{1+\varkappa^{2}}\ \widetilde{J}_{+}^{\alpha (2)},\\
M^{\alpha}_{-}&=J_{-}^{\alpha (0)}+\lambda\sqrt{1+\varkappa^{2}}\ J_{-}^{\alpha (2)},
\end{align}
where $\lambda$ is the spectral parameter. In terms of the vector:
$$\mathcal{L}_{\alpha}=L_{+\alpha}+M_{-\alpha},$$
the zero-curvature condition of \cite{Delduc:2013qra} was given by:
\begin{equation}
 \partial_{\alpha}\mathcal{L}_{\beta}-\partial_{\beta}\mathcal{L}_{\alpha}+\left[\mathcal{L}_{\alpha},\mathcal{L}_{\beta}\right]=0.
\label{zerocurvatureetadeformations}
\end{equation}
For the spinning string solution of equation \eqref{spininganzats} we have that $\partial_{\tau}\mathcal{L}_{\sigma}=0$, which is due to the fact that there is no $\tau$-dependence in the fields. Therefore, for spinning strings in $\eta$-deformed $\ads$ the zero-curvature condition \eqref{zerocurvatureetadeformations} reduces to:
\begin{equation}
\partial_{\sigma}\mathcal{L}_{\tau}=\left[\mathcal{L}_{\tau},\mathcal{L}_{\sigma}\right].
\label{zerocurvaturespinning}
\end{equation}
Using the $\mathcal{L}_{\tau}$ and $\mathcal{L}_{\sigma}$ from the spinning string ansatz, which are constructed as mentioned above, it is easy to check that equation \eqref{zerocurvaturespinning} follows from the equations of motion of the Lagrangian of equation \eqref{Lagrangianrxi}. The existence of this Lax pair implies that we can use $\mathcal{L}_{\tau}$ to generate a tower of integrals of motion, as we will see now.

\section{Deformed Uhlenbeck Integrals from Lax formalism}
\label{section5}
Using the Lax formalism we can generate integrals of motion by using:
\begin{equation}
Q_{k}(\lambda)=\text{Tr}\left[\left(\mathcal{L}_{\tau}\right)^{k}\right],
\end{equation}

\noindent where for every $k\in\mathbb{N}$ each coefficient in the power expansion of $Q_{k}(\lambda)$ in the spectral parameter $\lambda$ is a conserved quantity. Due to Newton's identities for the trace and the dimensions of the matrix $\mathcal{L}_{\tau}$, it is clear that for $k>4$ the conserved quantities obtained in this way can be written in terms of the conserved quantities resulting from the $Q_{k}(\lambda)$ with $k\in\{1,2,3,4\}$.\\

It is also clear that the Hamiltonian corresponding to the Lagrangian of equation \eqref{Lagrangianrxi} (which will be denoted by $\widetilde{H}$) is a conserved quantity since it does not have an explicit dependence on $\sigma$. Thus, it will be natural for $\widetilde{H}$ to appear among the conserved quantities obtained from the $Q_{k}$'s. In fact, by direct evaluation of each one of these conserved quantities, it can be shown that the only non-trivial conserved quantities that are independent of the Hamiltonian are $\widetilde{Q}_{1}$ and $\widetilde{Q}_{2}$, obtained from:
\begin{align}
\frac{1}{2!}\frac{d^{2}(\lambda^{4}Q_{4})}{d\lambda^{2}}\bigg|_{\lambda=0}&=-\frac{1}{4}(1-\varkappa^{4})\widetilde{H}^2+\widetilde{Q}_{1},\\
\frac{1}{4!}\frac{d^{4}(\lambda^{4}Q_{4})}{d\lambda^{4}}\bigg|_{\lambda=0}&=\frac{3-2\varkappa^{2}+3\varkappa^{4}}{8}\widetilde{H}^{2}+\widetilde{Q}_{2}.
\end{align}
These conserved quantities $\widetilde{Q}_{1}$ and $\widetilde{Q}_{2}$ are very large expressions which go up to orders $(\pi_{r})^2$ and $(\pi_{\xi})^{4}$; being $\pi_{r}$ and $\pi_{\xi}$ the canonical momentum conjugated to coordinates $r$ and $\xi$, respectively. In contrast, for the Neumann model it can be shown that in these $(r,\xi)$ coordinates the integrals $F_{i}$ only go up to orders $(\pi_{r})^2$ and $(\pi_{\xi})^{2}$ in momenta. Moreover, in the undeformed limit ($\varkappa\rightarrow 0$) the conserved quantities $\widetilde{Q}_{1}$ and $\widetilde{Q}_{2}$ reduce to linear combinations of $F_{1}$ and $F_{2}$, plus a constant. This behaviour suggests the existence of deformed conserved quantities ${\mathscr F}_{i}$ satisfying $\lim_{\varkappa\rightarrow 0}{\mathscr F}_{i}=F_{i}$, which will be up to orders $(\pi_{r})^2$ and $(\pi_{\xi})^{4}$ in momenta. These deformed integrals ${\mathscr F}_{i}$ will play the role of the Uhlenbeck integrals for the deformed system.\\

Based on the behaviour of $\widetilde{Q}_{1}$ and $\widetilde{Q}_{2}$ in the undeformed limit and the fact that in the undeformed case we only have 2 independent quantities $F_{i}$; we would expect the conserved quantities $\widetilde{H}$, $\widetilde{Q}_{1}$ and $\widetilde{Q}_{2}$ to be a linear combination of the ${\mathscr F}_{i}$ in the deformed case. This would correspond to having:
\begin{align}
\widetilde{H}=&A_{1}{\mathscr F}_{1}+A_{2}{\mathscr F}_{2}+A_{3}\label{systemeq1},\\
\widetilde{Q}_{1}=&B_{1}{\mathscr F}_{1}+B_{2}{\mathscr F}_{2}+B_{3}\label{systemeq2},\\
\widetilde{Q}_{2}=&C_{1}{\mathscr F}_{1}+C_{2}{\mathscr F}_{2}+C_{3}\label{systemeq3},
\end{align}
with $A_{1}$, $A_{2}$, $A_{3}$, $B_{1}$, $B_{2}$, $B_{3}$, $C_{1}$, $C_{2}$ and $C_{3}$ unknown coefficients, which can in principle depend on the deformation parameter $\varkappa$. This system of 3 equations can be treated as 2 systems of 2 equations each: The first one corresponds to equations \eqref{systemeq1} and \eqref{systemeq2}, while the second one corresponds to equations  \eqref{systemeq1} and \eqref{systemeq3}. Since we know the left hand side of these 3 equations we can solve the 2 systems separately: From the first system we can solve for ${\mathscr F}_{1}(A_{1},A_{2},A_{3},B_{1},B_{2},B_{3})$ and ${\mathscr F}_{2}(A_{1},A_{2},A_{3},B_{1},B_{2},B_{3})$, while for the second system we can solve for ${\mathscr F}_{1}(A_{1},A_{2},A_{3},C_{1},C_{2},C_{3})$ and ${\mathscr F}_{2}(A_{1},A_{2},A_{3},C_{1},C_{2},C_{3})$. The ${\mathscr F}_{1}$ and ${\mathscr F}_{2}$ we are looking for should be simultaneously the solution to these 2 systems, thus the 2 solutions for ${\mathscr F}_{1}$ and the 2 solutions for ${\mathscr F}_{2}$ should be equal. This means that we can match the solutions power by power in $r'$ and $\xi'$ (or equivalently in $\pi_{r}$ and $\pi_{\xi}$). By doing this we can solve for the coefficients $C_{i}$ obtaining:
$$C_{1}=\frac{2 B_{1} \left(\varkappa ^2-1\right)}{\varkappa ^2+1},\ \ \ \ \ \ \ \ \ \  C_{2}=\frac{2 B_{2} \left(\varkappa ^2-1\right)}{\varkappa ^2+1},$$
$$C_{3}=\frac{8 B_{3} \left(\varkappa ^2-1\right)+\left(\varkappa ^2+1\right) \left(\omega_1^4+6\omega_1^2 \left(\omega_2^2+\omega_3^2\right)+\omega_2^4+6 \omega_2^2 \omega_3^2+\omega_3^4\right)}{4 \left(\varkappa ^2+1\right)}.$$

It can be checked that the two ${\mathscr F}_{i}(A_{1},A_{2},A_{3},B_{1},B_{2},B_{3})$ obtained in this way satisfy $\{{\mathscr F}_{1},{\mathscr F}_{2}\}=0$, and by construction also satisfy $\{\widetilde{H},{\mathscr F}_{i}\}=0$ with $i=1,2$ (using canonical Poisson brackets). \\
For the undeformed case it can be checked that once the 3 expressions for the $F_{i}$ are substituted in the Hamiltonian (recall equation \eqref{undeformedhamiltonianF}), we have that:
\begin{equation}
\frac{\partial^{2} H}{ {\partial\omega_{i}}^{2}}=F_{i}\bigg|_{\pi_{r},\pi_{\xi}=0}\ \ \ \ \ \ \forall i\in\{1,2,3\}.
\end{equation}
Using the deformed Hamiltonian $\widetilde{H}$ (which was obtained from \eqref{Lagrangianrxi}) it is easy to check that $\partial^{2} \widetilde{H}/ {\partial\omega_{i}}^{2}=\partial^{2} H/ {\partial\omega_{i}}^{2}$. Since we want our deformed ${\mathscr F}_{i}$ to coincide with the Uhlenbeck integrals when $\varkappa\rightarrow0$, we will impose for the deformed case that:
\begin{equation}
\frac{\partial^{2} \widetilde{H}}{ {\partial\omega_{i}}^{2}}={\mathscr F}_{i}\bigg|_{\pi_{r},\pi_{\xi}=0}\ \ \ \ \ \ \forall i\in\{1,2\}\ ,
\label{zouk}
\end{equation}
which is equivalent to imposing the condition ${\mathscr F}_{i}|_{\pi_{r},\pi_{\xi}=0}=F_{i}|_{\pi_{r},\pi_{\xi}=0}$. We can then match both sides of equation \eqref{zouk} by comparing them term by term in powers of $r$. Doing this first for $i=1$ one finds the coefficients $B_{i}$ in terms of the $A_{i}$ coefficients:
\begin{align*}
B_{1}&=-\frac{1}{4} \left(1+\varkappa ^2\right) \left[\left(\omega_1^2-\omega_2^2\right) \left(\omega_1^2-\omega_3^2\right)-A_{1} \left(3 \omega_1^2+\omega_2^2+\omega_3^2\right)\right],\\
B_{2}&=\frac{1}{4} A_{2} \left(1+\varkappa ^2\right) \left(3\omega_1^2+\omega_2^2+\omega_3^2\right),\\
B_{3}&=\frac{1}{4} \left(1+\varkappa ^2\right) \left[A_{3} \left(3 \omega_1^2+\omega_2^2+\omega_3^2\right)+\omega_2^2 \omega_3^2\right].
\end{align*}
Repeating this procedure for $i=2$ allows us to find the coefficients $A_{i}$, and consequently fix the coefficients $B_{i}$ and $C_{i}$ found previously. The final expressions for all of the coefficients are given by:
\begin{align*}
A_{1}&=\frac{1}{2}(\omega_1^2-\omega_3^2),\ \ \ \ \ \ \ \ \ \ A_{2}=\frac{1}{2}(\omega_2^2-\omega_3^2),\ \ \ \ \ \ \ \ \ \ \ \ \ A_{3}=\frac{1}{2}\omega_3^2,\\
B_{1}=&\frac{1}{8}(\omega_1^2+3\omega_2^2+\omega_3^2)(1+\varkappa^{2})(\omega_1^2-\omega_3^2),\\
B_{2}=&\frac{1}{8}(3\omega_1^2+\omega_2^2+\omega_3^2)(1+\varkappa^{2})(\omega_2^2-\omega_3^2),\\
B_{3}=&\frac{1}{8}\omega_3^2 (3\omega_1^2+3\omega_2^2+\omega_3^2)(1+\varkappa^{2}),\\
C_{1}=&-\frac{1}{4}(\omega_1^2+3\omega_2^2+\omega_3^2)(1-\varkappa^{2})(\omega_1^2-\omega_3^2),\\
C_{2}=&-\frac{1}{4}(3\omega_1^2+\omega_2^2+\omega_3^2)(1-\varkappa^{2})(\omega_2^2-\omega_3^2),\\
C_{3}=&\frac{1}{4}(\omega_1^4+\omega_2^4+\omega_3^4 \varkappa^{2}+3\omega_2^2 \omega_3^2(1+\varkappa^{2})+3\omega_1^2(2 \omega_2^2+\omega_3^2(1+\varkappa^{2}))).
\end{align*}
From these coefficients we get the expressions for ${\mathscr F}_{1}$ and ${\mathscr F}_{2}$. Equation \eqref{systemeq1} and the results for the coefficients $A_{i}$ suggest that we can introduce ${\mathscr F}_{3}$ by doing:
\begin{equation}
1={\mathscr F}_{1}+{\mathscr F}_{2}+{\mathscr F}_{3}.
\label{deformedsumF}
\end{equation}
Then, the Hamiltonian for the deformed model is given by:
\begin{equation}
\widetilde{H}=\frac{1}{2}\sum_{i=1}^{3}\omega_i^2{\mathscr F}_{i}.
\label{hamiltoniandeformedF}
\end{equation}

Equations \eqref{deformedsumF} and \eqref{hamiltoniandeformedF} are the deformed versions of equations \eqref{undeformedFproperties} and \eqref{undeformedhamiltonianF} of the undeformed Neumann model. The full expressions for the ${\mathscr F}_{i}$ found by this method are given in the Appendix \ref{apendice}.\\
It can be checked that, using the canonical Poisson bracket $\{\pi_{r},r\}=1$ and $\{\pi_{\xi},\xi\}=1$, the three conserved quantities  ${\mathscr F}_{i}$ satisfy:
\begin{equation}
\{\widetilde{H}, {\mathscr F}_{i}\}=0 \ \ \ \forall i \in\{1,2,3\}, \ \ \ \ \ \ \ \ \ \ \ \ \ \{ {\mathscr F}_{i}, {\mathscr F}_{j}\}=0 \ \ \ \ \forall  i,j\in\{1,2,3\}.
\end{equation}

\section{Dirac Formulation for the Deformed Neumann Model}
\label{section6}
So far we have used coordinates $(r,\xi)$ and their momenta $(\pi_{r},\pi_{\xi})$, this allowed us to work in a 4-dimensional phase space without constraints, where we used canonical Poisson brackets. As was mentioned earlier, for $N=3$ the Neumann model is usually presented in terms of Dirac brackets $\{ , \}_{D}$ embedded in a 6-dimensional phase space corresponding to coordinates $x_{i}$ and their momenta $\pi_{i}$ with $i\in\{1,2,3\}$, but subject to the 2 constraints of equation \eqref{constrainsundeformed}.\\
In order to perform the transition to this second formalism, we first move to coordinates $x_{1}$ and $x_{2}$ by using:
\begin{equation}
x_{1}=r\cos\xi,\ \ \ \ \ \ \ \ \ \ \ \ \ \ x_{2}=r\sin\xi.
\label{addition1}
\end{equation}
Conjugated to these two coordinates we will have momenta $p_{1}$ and $p_{2}$, respectively. By doing this change of coordinates we have done the transition from phase space coordinates $(r,\xi,\pi_{r},\pi_{\xi})$ to $(x_{1},x_{2},p_{1},p_{2})$. Now, we will proceed to increase the dimensionality of the phase space by introducing a new coordinate and momentum.

The third coordinate is introduced by making use of:
\begin{equation}
\sum_{i=1}^{3}x_i^2=1 .
\label{constrain1deformed}
\end{equation}
We now have to introduce a third momentum associated with coordinate $x_{3}$, this is done by means of a transformation from $(p_{1},p_{2})$ to new momenta $(\pi_{1},\pi_{2},\pi_{3})$. The transformation we will use is given by:
\begin{equation}
p_{1}\rightarrow-\frac{\pi_{3}x_{1}-\pi_{1}x_{3}}{x_{3}},\ \ \ \ p_{2}\rightarrow-\frac{\pi_{3}x_{2}-\pi_{2}x_{3}}{x_{3}}.
\end{equation}

Using this transformation we can write the deformed integrals in a more compact form:
\begin{align}
\label{Ffinal1}
 {\mathscr F}_{1}=&F_{1}+\frac{{\sum\limits_{i = 1}^4 {{n_i}} {\varkappa ^i}}}{{\left( {\omega _1^2 - \omega _2^2} \right)\left( {\omega _1^2 - \omega _3^2} \right)}} - \frac{{2\varkappa {J_{13}}{x_1}{x_3}{\omega _3}}}{{\omega _1^2 - \omega _3^2}}- \frac{{2\varkappa {J_{12}}{x_1}{x_2}{\omega _1}}}{{\omega _1^2 - \omega _2^2}}\\
&+ \frac{{{\varkappa ^2}J_{12}^2x_2^2}}{{\omega _1^2 - \omega _2^2}}+\frac{{{\varkappa ^2}J_{13}^2(x_1^2+x_2^2)}}{{\omega _1^2 - \omega _3^2}} - \frac{{{\varkappa ^2}J_{12}^2x_3^2\omega _1^2}}{{\left( {\omega _1^2 - \omega _2^2} \right)\left( {\omega _1^2 - \omega _3^2} \right)}} \ ,\nonumber
\end{align}
\begin{align}
\label{Ffinal2}
 {\mathscr F}_{2}=&F_{2}+\frac{{\sum\limits_{i = 1}^4 {{n_i}{\varkappa ^i}} }}{{\left( {\omega _2^2 - \omega _1^2} \right)\left( {\omega _2^2 - \omega _3^2} \right)}} - \frac{{2\varkappa {J_{23}}{x_2}{x_3}{\omega _3}}}{{\omega _2^2 - \omega _3^2}} - \frac{{2\varkappa {J_{12}}{x_1}{x_2}{\omega _1}}}{{\omega _2^2 - \omega _1^2}}\\
&+ \frac{{{\varkappa ^2}J_{12}^2x_2^2}}{{\omega _2^2 - \omega _1^2}}+ \frac{{{\varkappa ^2}J_{23}^2(x_1^2+x_2^2)}}{{\omega _2^2 - \omega _3^2}} - \frac{{{\varkappa ^2}J_{12}^2x_3^2\omega _2^2}}{{\left( {\omega _2^2 - \omega _1^2} \right)\left( {\omega _2^2 - \omega _3^2} \right)}}\ ,\nonumber
\end{align}
\begin{align}
\label{Ffinal3}
 {\mathscr F}_{3}=&F_{3}+\frac{{\sum\limits_{i = 1}^4 {{n_i}{\varkappa ^i}} }}{{\left( {\omega _3^2 - \omega _1^2} \right)\left( {\omega _3^2 - \omega _2^2} \right)}} - \frac{{2\varkappa {J_{23}}{x_2}{x_3}{\omega _3}}}{{\omega _3^2 - \omega _2^2}} - \frac{{2\varkappa {J_{13}}{x_1}{x_3}{\omega _3}}}{{\omega _3^2 - \omega _1^2}}\\
 &+ \frac{{{\varkappa ^2}J_{13}^2(x_1^2+x_2^2)}}{{\omega _3^2 - \omega _1^2}} + \frac{{{\varkappa ^2}J_{23}^2(x_1^2+x_2^2)}}{{\omega _3^2 - \omega _2^2}} - \frac{{{\varkappa ^2}J_{12}^2x_3^2\omega _3^2}}{{\left( {\omega _3^2 - \omega _1^2} \right)\left( {\omega _3^2 - \omega _2^2} \right)}} \ ,\nonumber
\end{align}
where the $F_{i}$ are the Uhlenbeck integrals of the undeformed Neumann model (recall equation \eqref{undeformedFproperties}), while the $n_{i}$ are given by:
\begin{align*}
n_{1}=& - 2{J_{12}}{J_{13}}{J_{23}}{\omega _1}\ ,\\
n_{2}=&- J_{12}^2x_3^2\omega _3^2 + 2{J_{12}}{x_3}{\omega _1}{\omega _3}\left( {{J_{23}}{x_1} + {J_{13}}{x_2}} \right) - J_{12}^2J_{13}^2\ ,\\
n_{3}=&2{J_{12}}{J_{13}}\left[ {{J_{12}}{x_1}{x_3}{\omega _3} - {J_{23}}\left( {x_1^2 + x_2^2} \right){\omega _1}} \right]\ ,\\
n_{4}=& - J_{12}^2J_{13}^2\left( {x_1^2 + x_2^2} \right)\ .
\end{align*}

From the set of equations for the $ {\mathscr F}_{i}$ it is easily seen that in the limit $\varkappa\rightarrow0$ the expressions for the ${\mathscr F}_{i}$ reduce to the $F_{i}$ of the undeformed Neumann model.\\

The deformed Hamiltonian $\widetilde{H}$ written in terms of phase space coordinates $(x_{i},\pi_{i})$ is obtained by substituting equations \eqref{Ffinal1}, \eqref{Ffinal2} and \eqref{Ffinal3} in equation \eqref{hamiltoniandeformedF}, doing this one gets the following expression:
\begin{align}
{\widetilde H}=&\frac{1}{4}\sum_{i\neq j}J_{ij}^{2} +\frac{1}{2}\sum_{i=1}^{3}\omega_i^2 x_i^2 - \varkappa \left( {\omega_{1} {x_1}{x_2}{J_{12}} + \omega_{3} {x_1}{x_3}{J_{13}} + \omega_{3} {x_2}{x_3}{J_{23}}} \right)\nonumber\\
 &+ \frac{{{\varkappa ^2}}}{2}\left[ {\left( {x_2^2 - x_3^2} \right)J_{12}^2 + \left( {x_1^2 + x_2^2} \right)J_{13}^2 + \left( {x_1^2 + x_2^2} \right)J_{23}^2} \right]\ .
\label{grandeHHH}
\end{align}

The expression above clearly coincides with the one in equation \eqref{alternativeH} when taking the limit $\varkappa \rightarrow 0$.

As we saw before, the Neumann model consists of 2 constraints, being \eqref{constrain1deformed} the first of them. By construction the deformed model satisfies \eqref{constrain1deformed}, but we also want the deformed system to satisfy the second constraint, thus we will impose that the new momenta $\pi_{i}$ satisfy:
\begin{equation}
 \sum_{i}x_{i}\pi_{i}=0\ .
\label{constrain2deformed}
\end{equation}

One can check that the Dirac bracket constructed from phase space coordinates $(x_{i},\pi_{i})$ and constraints \eqref{constrain1deformed} and \eqref{constrain2deformed} is such that:
\begin{equation}
\{\widetilde{H}, {\mathscr F}_{i}\}_{D}=0\ ,\ \ \ \ \ \ \ \ \{{\mathscr F}_{i}, {\mathscr F}_{j}\}_{D}=0 \ \ \ \ \ \forall i,j \in \{1,2,3\}.
\end{equation}
The attentive reader may have noticed that we have introduced new momenta $\pi_{i}$, but so far we have not established their dependence in terms of the $x_{i}$ and ${\dot x}_{i}$. For this we will use the constraint \eqref{constrain2deformed} and the equations:
\begin{align}
{\dot x_{1}}&=\{\widetilde{H},x_{1} \}_{D}\label{esoeq1}\ ,\\
{\dot x_{2}}&=\{\widetilde{H},x_{2} \}_{D}\label{esoeq2}\ .
\end{align}
By evaluating the right hand side of \eqref{esoeq1} and \eqref{esoeq2}, we are left with a system of 3 equations relating $\pi_{i}$, $x_{i}$ and ${\dot x}_{i}$. Solving for the momenta $\pi_{i}$ one finds:
\begin{align*}
\pi_{1}=&\frac{1}{u}\left[{{\dot x}_1} - \varkappa {x_1}\left( {x_2^2{\omega _1} + x_3^2{\omega _3}} \right) + {\varkappa ^2}{x_2}\left( {{x_2}{{\dot x}_1}\left( {1 + x_1^2} \right) - {x_1}{{\dot x}_2}\left( {1 - x_2^2} \right)} \right) \right. \\
& \left.	- {\varkappa ^3}{x_1}x_2^2\left( {x_1^2 + x_2^2} \right)\left( {{\omega _1} + x_3^2{\omega _3}} \right)\right],\\
\pi_{2}=&\frac{1}{u}\left[{{\dot x}_2} - \varkappa {x_2}\left( {x_3^2{\omega _3} - x_1^2{\omega _1}} \right) + {\varkappa ^2}\left( {\left( {x_1^2 + x_2^4} \right){{\dot x}_2} - {x_1}{x_2}\left( {1 - x_2^2} \right){{\dot x}_1}} \right)\right.\\
& \left. + {\varkappa ^3}{x_2}\left( {x_1^2 + x_2^2} \right)\left( {x_1^2{\omega _1} - x_3^2x_2^2{\omega _3}} \right)\right] ,\\
\pi_{3}=&\frac{{{{\dot x}_3} + \varkappa {\omega _3}{x_3}\left( {x_1^2 + x_2^2} \right)}}{{1 + {\varkappa ^2}\left( {x_1^2 + x_2^2} \right)}}\ ,
\end{align*}
where:
$$u={\left( {1 + {\varkappa ^2}\left( {x_1^2 + x_2^2} \right)} \right)\left( {1 + {\varkappa ^2}x_2^2\left( {x_1^2 + x_2^2} \right)} \right)}.$$
Once again, in the undeformed limit we recover the results for the Neumann model, namely, $\pi_{i}={\dot x}_{i}$.\\

Having constructed the momenta $\pi_{i}$ in this way, one can check that indeed the Dirac bracket will define the time evolution of the system. In other words, that:
\begin{equation}
{\dot x_{i}}=\{\widetilde{H},x_{i} \}_{D}\ ,\ \ \ \ \ \ \ {\dot \pi_{i}}=\{\widetilde{H},\pi_{i} \}_{D}\ .
\end{equation}
The equation from the left for the cases of $i=1, 2$ is satisfied by construction, and for the case of $i=3$ it is  verified by making use of $x_{3}=\sqrt{1-x_{1}^2-x_{2}^2}$. Meanwhile, the equation on the right can be verified by using the expressions for the $\pi_{i}$, the constraint \eqref{constrain1deformed} and the Euler-Lagrange equations of motion.\\

Let us briefly summarize the main results of the present section. Starting from the unconstrained formulation in terms of $(r,\xi)$ coordinates we found a formulation for an integrable deformation of the Neumann model in terms of coordinates $x_{i}$ and their momenta $\pi_{i}$. In this formulation the system is given by the Hamiltonian \eqref{grandeHHH} subject to the constraints \eqref{constrain1deformed} and \eqref{constrain2deformed}, and its evolution is determined in terms of Dirac brackets. The integrals of motion that guarantee the explicit Liouville integrability of the model are given by \eqref{Ffinal1}, \eqref{Ffinal2} and \eqref{Ffinal3}.

\section{Conclusions}
In this work we have found a new integrable model by studying bosonic spinning strings on $\eta$-deformed $\ads$. This model corresponds to a highly non-trivial deformation of the celebrated Neumann model. The Liouville integrability of the model was made explicit by finding the deformed integrals of motion, which are the equivalent of the Uhlenbeck integrals of the undeformed model. Naturally, this being a novel integrable model, there are many open questions that have yet to be addressed.

As is well known, practically all known integrable systems admit a Lax representation from which a tower of conserved quantities can be generated, thus guaranteeing the integrability of the model. Throughout the derivation of the results found in this paper for the deformed Neumann model, the $4\times4$ Lax formalism inherited from the spinning strings ansatz in $\eta$-deformed $\ads$ played a crucial role. However, its structure is not very transparent and it is characterized by large expressions in each of its components. For the undeformed Neumann model relatively simple Lax pair constructions in terms of $2\times2$ and $3\times 3$ matrices have been found \cite{Avan:1991ib,Avan:1989dn}. These formulations were obtained by considering the Neumann model as the result of a 
reduction by symmetry of another integrable system with a larger unconstrained phase space \cite{BabelonBernardTalon,Avan:1989dn,JMoser}. This other system corresponds to the Hamiltonian \eqref{alternativeH} with unconstrained phase space coordinates $(x_{i},\pi_{i})$, and with time evolution determined by canonical Poisson brackets. For the deformed Neumann model proposed in this paper perhaps similar Lax constructions can be formulated. In that case the most likely starting point would be to consider the Hamiltonian \eqref{grandeHHH}, but using canonical Poisson brackets instead of Dirac brackets, thus resulting in a higher dimensional phase space since the constraints \eqref{constrain1deformed} and \eqref{constrain2deformed} are dropped. It is not yet clear how these $2\times2$ and $3\times3$ Lax structures can be realized for the deformed Neumann model proposed here, but this is clearly a very interesting question for which further work is required.

The work of J. Moser showed that the Neumann model also describes the geodesic motion on a ellipsoid \cite{BabelonBernardTalon}. The latter was shown to be integrable by Jacobi, who solved the problem by introducing ellipsoidal coordinates and employing the method of separation of variables. This appropriate choice of coordinates also allows for the separation of variables of the Neumann model, as was shown for the $N=3$ case by C. Neumann himself \cite{BabelonBernardTalon}. For the deformed Neumann model presented here, this particular set of coordinates does not allow for the separation of variables. Finding an adequate set of coordinates is therefore one more important and interesting challenge.

Another open question would be to generalize the results of this paper to $N>3$, since the results found here correspond to a deformation of the $N=3$ Neumann model. This task is highly non-trivial due to the manifest asymmetry in the coordinates $x_{i}$. For the case of generic $N$, one should start from the sigma-model on the coset space ${\rm SO}(N+1)/{\rm SO}(N)$ equivalent to ${\rm S}^{N}$, just as was the case for the five-sphere and the coset ${\rm SO}(6)/{\rm SO}(5)$. Then, one would need to deform this model using the same construction as in the case of the five-sphere, and finally perform a reduction on the spinning string ansatz. A procedure like the one proposed here should in principle provide a deformation for the Neumann model with generic $N$.

An interesting problem would be to introduce a generalized spinning string ansatz in the same spirit as the one proposed in \cite{Arutyunov:2003za}. For undeformed $\ads$ it was shown in \cite {Arutyunov:2003za} that the introduction of this ansatz allows for the description of rotating strings in terms of the $N=3$ Neumann-Rosochatius integrable system. The integrability of the $N=3$ Neumann-Rosochatius model is a consequence of the fact that it is equivalent to a special case of the $N=6$ Neumann model. By introducing a generalized rotating ansatz in $\eta$-deformed $\ads$, one would expect to find a deformation of the Neumann-Rosochatius system. However, proving its Liouville integrability would be highly non-trivial due to the difficulty of finding the corresponding integrals of motion. One can only conjecture that, in analogy with the undeformed model, these integrals of motion are somehow related to the ones of an $\eta$-deformed $N=6$ Neumann model.

One could also consider an even more generic string solution like the one presented in \cite{Kruczenski:2006pk}, where spiky strings and magnon solutions are obtained by means of a generalized Neumann-Rosochatius ansatz. For the $\ads$ background, the Lagrangian of such a system is shown to be described by a Neumann system plus a magnetic field interaction \cite{Kruczenski:2006pk}. Thus, for the study of giant magnons in the case of the $\eta$-deformed background \cite{Arutynov:2014ota,Khouchen:2014kaa,Ahn:2014aqa}, systems like the one proposed in this article might prove useful.

\section*{Acknowledgements}

We would like to thank Riccardo Borsato for discussions, and Wellington Galleas, Jules Lamers, Arkady Tseytlin and Stijn van Tongeren for useful comments on the manuscript. G.A. acknowledges the support by the Netherlands Organization for Scientific Research (NWO) under the VICI grant 680-47-602. The work by G.A. is also a part of the ERC Advanced grant research programme No. 246974,  {\it ``Supersymmetry: a window to non-perturbative physics"} and of the D-ITP consortium, a program of the NWO that is funded by the Dutch Ministry of Education, Culture and Science (OCW). The work of D.M. was supported in part by Compa\~n\'ia Colombiana Automotriz S.A. and Fundaci\'on Mazda para el Arte y la Ciencia.

\appendix 

\section{Geodesic Motion}\label{apendice}
In this section we will briefly comment on geodesic motion on the $\eta$-deformed background and explain that it is also governed by an integrable model.  As in the main text, without loss of generality, we will be interested in the most general solution for particle-like strings in the center of $\rm{AdS}$. For this class of solutions the fields $r$, $\xi$, $\phi$, $\phi_{1}$ and $\phi_{2}$ will depend only on the world-sheet time $\tau$. This type of solution is described by the corresponding Lagrangian obtained from equation \eqref{L} by dropping the $\sigma$-dependence, and its Hamiltonian is given by:
\begin{align}
\widetilde{H}&= \frac{1}{2}\bigg[ \pi _r^2\left( {1 - {r^2}} \right)\left( {1 + {\varkappa ^2}{r^2}} \right) + \frac{{\pi _\xi ^2\left( {1 + {\varkappa ^2}{r^4}{{\sin }^2}\xi } \right)}}{{{r^2}}} + \frac{{\pi _{\phi_{\rm{1}}}^2\left( {1 + {\varkappa ^2}{r^4}{{\sin }^2}\xi } \right)}}{{{r^2}{{\cos }^2}\xi }}\label{HGeodesics} \\
&\quad \quad + \frac{{\pi _{\phi_{\rm{2}}}^2}}{{{r^2}{{\sin }^2}\xi }} + \frac{{\pi _{\phi}^2\left( {1 + {\varkappa ^2}{r^2}} \right)}}{{1 - {r^2}}} \bigg]\ ,\nonumber
\end{align}
where the canonical momenta are:
\begin{align}
{\pi _r} = \frac{{\dot r}}{{\left( {1 - {r^2}} \right)\left( {1 + {\varkappa ^2}{r^2}} \right)}},\quad\quad\quad\quad {\pi _\xi } = \frac{{{r^2}\dot \xi }}{{1 + {\varkappa ^2}{r^4}{{\sin }^2}\xi }},\quad\quad\quad\label{momentaGeo}\\
 {\pi _{\phi_{\rm{1}}}} = \frac{{{r^2}{{\cos }^2}\xi \mathop {{\phi _1}}\limits^. }}{{1 + {\varkappa ^2}{r^4}{{\sin }^2}\xi }},\quad\quad{\pi _{\phi_{\rm{2}}}} = {r^2}\mathop {{\phi _2}}\limits^. {\sin ^2}\xi ,\quad\quad{\pi _\phi } = \frac{{\left( {1 - {r^2}} \right)\dot \phi }}{{1 + {\varkappa ^2}{r^2}}}\ ,\nonumber
\end{align}
with a dot indicating a derivative with respect to $\tau$. Since the Hamiltonian has no explicit $\tau$ dependence, it is an integral of motion. Additionally, this system has other three commuting integrals  originating from the isometries in the angles $\phi_{1}$, $\phi_{2}$ and $\phi$:
\begin{equation}
{\widetilde J_1} = {\pi _{{\phi _1}}},\quad\quad {\widetilde J_2} = {\pi _{{\phi _2}}},\quad\quad {\widetilde J_3} = {\pi _{\phi}}\ .
\label{jsgeodesics}
\end{equation}
The Hamiltonian \eqref{HGeodesics} can easily be written in terms of the three $\widetilde{J}_{i}$, the coordinates $r$, $\xi$ and their respective momenta $\pi_{r}$ and $\pi_{\xi}$. By writing the later 2 momenta in terms of the derivatives of $r$ and $\xi$, and then moving to the coordinates $x_{i}$ introduced in equations \eqref{addition1} and \eqref{constrain1deformed}, it is possible to rewrite $\widetilde{H}$ in the following way:

\begin{align}
\widetilde{H}=\frac{1}{2}\left[ \frac{{\dot x_1^2\left( {1 + {\varkappa ^2}\left( {1 - x_3^2} \right)} \right) + \dot x_2^2\left( {1 + {\varkappa ^2}\left( {1 - x_3^2} \right)} \right) + \dot x_3^2\left( {1 + {\varkappa ^2}\left( {x_2^2 - x_3^2} \right)} \right)}}{{\left( {1 + {\varkappa ^2}\left( {1 - x_3^2} \right)} \right)\left( {1 + {\varkappa ^2}x_2^2\left( {1 - x_3^2} \right)} \right)}}\right.\label{HamGeox}\\
+\left. \frac{{{{\tilde J}_1}^2\left( {1 + {\varkappa ^2}x_2^2\left( {1 - x_3^2} \right)} \right)}}{{x_1^2}} + \frac{{{{\tilde J}_2}^2}}{{x_2^2}} + \frac{{{{\tilde J}_3}^2\left( {1 + {\varkappa ^2}\left( {1 - x_3^2} \right)} \right)}}{{x_3^2}} \right]\ ,\nonumber
\end{align}
subject to the constraint $\sum_{i=1}^{3}x_{i}^{2}=1$. In this expression the $\widetilde{J}_{i}$ play the role of fixed constants, while the dynamics are described in terms of the coordinates $x_{i}$ and their derivatives.\\

\subsection{Geodesics on $\ads$}

In order to get some insight into the system described by this Hamiltonian, we will first study the undeformed limit of $\varkappa\rightarrow 0$. For undeformed $\ads$ this type of solutions have been treated in a similar manner in \cite{Frolov:2005iq}; here we intend to expand on this discussion by giving the explicit expressions for an adequate set of integrals of motion, and thus put this system in the framework of Liouville integrability. We will use a tilde to denote the deformed expressions $\widetilde{H}$ and $\widetilde{J}_{i}$, while $H$ and $J_{i}$ represent their undeformed counterparts. \\

In the undeformed limit equation \eqref{HamGeox} reduces to:
\begin{equation}
H = {\lim _{\kappa  \to 0}} \, \tilde H = \frac{1}{2}\left[ {\pi_1^2 + \pi_2^2 + \pi_3^2 + \frac{{J_1^2}}{{x_1^2}} + \frac{{J_2^2}}{{x_2^2}} + \frac{{J_3^2}}{{x_3^2}}} \right],
\end{equation}
where $\pi_{i}=\dot x_{i}$ denotes the canonical momentum conjugated to $x_{i}$. The system governed by $H$ and subject to constraints
\eqref{constrain1deformed} and \eqref{constrain2deformed}, is nothing else but the Rosochatius integrable system \cite{RosochatiusAA,BorisovAA1,BorisovAA2}. This integrable model has the following independent integrals of motion:
\begin{equation}{F_{ij}} = J_{ij}^2 + \frac{{J_i^2x_j^2}}{{x_i^2}} + \frac{{J_j^2x_i^2}}{{x_j^2}}\ , \quad\quad \text{with}\  i,j\in\{1,2,3\}\ \text{and}\ i\neq j \ ,
\end{equation}
where $J_{ij}=x_{i}\pi_{j}-x_{j}\pi_{i}$. These 3 integrals have non-vanishing Dirac brackets between themselves, but the Hamiltonian along with one of them are enough to integrate the system; as a consequence the Rosochatius system is super-integrable.\\
\indent Using equations \eqref{addition1} and \eqref{constrain1deformed}, the three integrals of motion $F_{ij}$ can be written in terms of the angular coordinates of $\rm{S}^{5}$ and their respective angular momenta:
\begin{align}
{F_{12}} &= \pi _\xi ^2 + \pi _{\phi_{\rm{1}}}^2{\tan ^2}\xi  + \pi _{\phi_{\rm{2}}}^2{\cot ^2}\xi\ ,\label{integralroso1}\\
{F_{13}} &= \pi _r^2\left( {1 - {r^2}} \right){\cos ^2}\xi  + \frac{{\pi _\xi ^2\left( {1 - {r^2}} \right){{\sin }^2}\xi }}{{{r^2}}} - \frac{{{\pi _r}{\pi _\xi }\left( {1 - {r^2}} \right)\sin 2\xi }}{r}\label{integralroso2}\\
 &+ \frac{{\pi _{{\phi _1}}^2\left( {1 - {r^2}} \right){{\sec }^2}\xi }}{{{r^2}}} + \frac{{\pi _{{\phi}}^2{r^2}{{\cos }^2}\xi }}{{1 - {r^2}}}\ ,\nonumber\\
 {F_{23}} &= \pi _r^2\left( {1 - {r^2}} \right){\sin ^2}\xi  + \frac{{\pi _\xi ^2\left( {1 - {r^2}} \right){{\cos }^2}\xi }}{{{r^2}}} + \frac{{{\pi _r }{\pi _\xi}\left( {1 - {r^2}} \right)\sin 2\xi }}{r}\label{integralroso3}\\
 & + \frac{{\pi _{{\phi _2}}^2\left( {1 - {r^2}} \right){{\csc }^2}\xi }}{{{r^2}}} + \frac{{\pi _{{\phi}}^2{r^2}{{\sin }^2}\xi }}{{1 - {r^2}}}\ ,\nonumber
\end{align}
where in these expressions the momenta correspond to the undeformed limit of the ones introduced in equation \eqref{momentaGeo}. 

Hence, any of the $F_{ij}$ introduced above, along with the 3 integrals $J_{i}$ and the Hamiltonian $H$, form a set of 5 integrals in involution under the canonical Poisson brackets in angular coordinates, rendering the model integrable in the Liouville sense. For explicit integration it is convenient to use $F_{12}$ as it directly leads to separation of variables. Thus, in the undeformed case 
the geodesic motion is described by  the Rosochatius integrable model. 

It is worth noting, however, that this is not the only way to solve the undeformed model. Another approach consists in starting from the Lagrangian of $\ads$ in the embedding coordinates $X_{M}$ and 
$Y_{N}$, where $X_{M}$ with $M\in\{1,...,6\}$ and $Y_{N}$ with $N\in\{0,...,5\}$, parametrize the sphere and $\rm{AdS}$ respectively. In these coordinates, the geodesic solutions we are interested in are described by:
$$X_{M}=X_{M}(\tau)\ ,\quad\quad Y_{5}+iY_{0}=e^{i\kappa t}, \quad\quad Y_{j}=0\quad\forall j\in\{1,...,4\}\ ,$$
while the effective Lagrangian is given by:
$$L = \frac{1}{2}{\dot X_M}{\dot X_M} + \Lambda \left( {{X_M}{X_M} - 1} \right)\ ,$$
where $\Lambda$ plays the role of a Lagrangian multiplier.  The (unconstrained) Hamiltonian for this system can be written as:
$$H=\frac{1}{2}\sum_{M=1}^{6}\Pi_{M}\Pi_{M}\ ,$$
subject to the constraints:
$$\sum_{M=1}^{6}X_{M}X_{M}=1\ ,\quad\quad\quad\sum_{M=1}^{6}X_{M}\Pi_{M}=0\ ,$$
where $\Pi_{M}=\dot{X}_{M}$ denotes the canonical momenta conjugated to $X_{M}$. This system has the following set of 5 integrals of motion in involution, which make the system completely integrable \cite{Integralspapersphere}:
\begin{equation}
{H_k} = {\sum\limits_{1 \le i < j \le k} {\left( {{X_i}{\Pi _j} - {X_j}{\Pi _i}} \right)} ^2},\quad \text{with}\  k \in \{ 2,...,6\}\ . \label{algebraicapproach}
\end{equation}
These integrals are  constructed by considering the chain of subalgebras $\mathfrak{so}(2)\subset\mathfrak{so}(3)\subset...\subset\mathfrak{so}(6)$, so that $H_{k}$ is a Casimir of $\mathfrak{so}(k)$, and, as a consequence, it commutes with all other $H_{j}$ with $j< k$. Moreover, the integrals $H_{k}$ are independent since each time $k$ is increased, new extra terms are included \cite{Integralspapersphere}. Writing these integrals in terms of angular coordinates by means of \cite{Arutyunov:2013ega}:
$${X_1} + i{X_2} = r\cos \xi {e^{i{\phi _1}}},\quad {X_3} + i{X_4} = r\sin \xi {e^{i{\phi _2}}},\quad {X_5} + i{X_6} = \sqrt {1 - {r^2}} {e^{i\phi }},$$
it is easy to check that $H_{2}=J_{1}^2$ and $H_{6}=2H$, while $H_{4}=F_{12}+J_{1}^{2}+J_{2}^{2}$. In particular, $H_{4}$ has a vanishing Poisson bracket with the $J_{i}$'s and $H$, and it leads to separation of variables.

\subsection{On integrability of geodesics on $(\ads)_{\eta}$}

Now we turn our attention to geodesic motion in the $\eta$-deformed case. In order to show explicit integrability of geodesics in the $(\rm{S}^{5})_{\eta}$ system, in the framework of Liouville's theorem, it is necessary to find a fifth integral with vanishing Poisson brackets with the deformed Hamiltonian $\widetilde{H}$ of equation \eqref{HGeodesics} and the three $\widetilde{J}_{i}$ of equations \eqref{jsgeodesics}. Finding this fifth integral of motion would make possible a solution of geodesic motion in this deformed background, and for instance examine the periodicity properties of the solutions. 
It is also clear that the Hamiltonian  \eqref{HamGeox} corresponds to a highly non-trivial deformation of the Rosochatius model, although there is no guarantee that this deformation remains super-integrable, {\it i.e.} that the analogues of the integrals $F_{ij}$ given by  \eqref{integralroso1}, \eqref{integralroso2} and \eqref{integralroso3} do exist.

A natural way to find a fifth integral is to carry our a procedure  similar to the one explained in sections \ref{section4} and \ref{section5}, where a Lax representation is constructed and used to generate integrals of motion. By following the same steps described in detail in section \ref{section4}, but this time only assuming $\tau$ dependence on the fields $r$, $\xi$, $\phi$, $\phi_{1}$ and $\phi_{2}$, one can build a $4\times4$ Lax representation for this system. Since all $\sigma$ dependence has been eliminated, the zero-curvature condition of equation \eqref{zerocurvatureetadeformations} reduces to:
$$\partial_{\tau}\mathcal{L}_{\sigma}=\left[\mathcal{L}_{\sigma},\mathcal{L}_{\tau}\right]\ ,$$
where now $\tau$ plays the role of time and $\mathcal{L}_{\sigma}$ can be used to generate a tower of integrals of motion:
$$Q_{k}(\lambda)=\text{Tr}[(\mathcal{L}_{\sigma})^{k}]\ .$$
Once again, for a given $k$ the integrals of motion correspond to the coefficients in front of each power of $\lambda$ in a power expansion in the spectral parameter $\lambda$. Due to Newton's identities for the trace it suffices to consider $k\le4$, since traces of higher powers of $\mathcal{L}_{\sigma}$ can be expressed in terms of the $Q_{k}$ with $k\in\{1,..,4\}$. By explicit calculation of all these quantities one can show that the only independent integral of motion $\widetilde{Q}$, which can not be expressed in terms of $\widetilde{H}$ and the three $\widetilde{J_{i}}$, is obtained as;
$${\left. {\frac{1}{{2!}}\frac{{{d^2}\left( {{\lambda ^4}{Q_4}} \right)}}{{d{\lambda ^2}}}} \right|_{\lambda  = 0}} =  - \frac{{1 - {\varkappa ^4}}}{4}{\tilde H^2} - \frac{{{\varkappa ^2}(1 + {\varkappa ^2})}}{4}\bigg[\tilde H\left( {{{\tilde J}_1}^2 + {{\tilde J}_2}^2 + {{\tilde J}_3}^2} \right) - {\tilde Q}\bigg] \ ,$$
where ${\tilde Q}$ is given by the following expression:
\begin{align}
\widetilde{Q}&=\left( {1 + {\varkappa ^2}{r^2}} \right)\left( {1 - {r^2}} \right)\bigg[ {\frac{{\pi _\xi ^4{{\sin }^2}\xi }}{{{r^2}}}}{ - \frac{{\pi _\xi ^3{\pi _r}\sin 2\xi }}{r}}{ + \pi _r^2\pi _\xi ^2{{\cos }^2}\xi }{ + \frac{{2\pi _\xi ^2\pi _{{\phi _1}}^2{{\tan }^2}\xi }}{{{r^2}}}}\label{expQQQ1scaled}\\
&{ + \frac{{\pi _\xi ^2\pi _{{\phi _2}}^2}}{{{r^2}}}}{ + \pi _r^2\pi _{{\phi _2}}^2{{\cot }^2}\xi }{ - \frac{{2{\pi _r}{\pi _\xi }\pi _{{\phi _1}}^2\tan \xi }}{r}}{ - \frac{{2{\pi _r}{\pi _\xi }\pi _{{\phi _2}}^2\cot \xi }}{r}}{ + \frac{{\pi _{{\phi _1}}^4{{\tan }^2}\xi\  {{\sec }^2}\xi }}{{{r^2}}}}\bigg]\nonumber\\
&+ \pi _{{\phi _1}}^2\pi _\phi ^2\left( {{\varkappa ^2} + {{\sin }^2}\xi } \right){\sec ^2}\xi + \frac{{\pi _{{\phi _2}}^2\pi _\phi ^2\left( {{{\cos }^2}\xi  + {\varkappa ^2}\left( {1 - {r^2}{{\sin }^2}\xi } \right)} \right){{\csc }^2}\xi }}{{1 - {r^2}}}\nonumber\\
 &+ \frac{{\left( {1 + {\varkappa ^2}} \right)\pi _\xi ^2\pi _\phi ^2\left( {1 - {r^2}{{\sin }^2}\xi } \right)}}{{1 - {r^2}}}+ \frac{{\pi _{{\phi _1}}^2\pi _{{\phi _2}}^2\left( {1 + {\varkappa ^2}{r^2} - {r^2}\left( {1 + {\varkappa ^2}{r^2}{{\sin }^2}\xi } \right)} \right){{\sec }^2}\xi }}{{{r^2}}}\nonumber\ .
\end{align}
Further, one can verify by explicit calculation that  $\widetilde{Q}$ is in involution with $\widetilde{H}$ and $\widetilde{J}_{i}$'s, which guarantees the Liouville integrability of geodesic motion in the deformed background. In the undeformed limit ${{\tilde Q}}$ has the following structure
$$\mathop {\lim }\limits_{\varkappa  \to 0} {{{\tilde Q}}} = {{F_{12}}{F_{13}} - J_1^2{F_{23}} + J_2^2{F_{13}} + J_3^2{F_{12}}}\ .$$
This completes our discussion of geodesic motion on the $\eta$-deformed background. We have not attempted here to separate the variables and obtain an explicit description of trajectories, as this is clearly a subject of a separate thorough investigation.

\section{Deformed Integrals in $r$ and $\xi$ Coordinates}\label{apendice}
In these coordinates the undeformed Uhlenbeck integrals can be written in the following way:
\begin{align}
F_{1}&=r^2 \cos ^2\xi+\frac{\pi _{\xi }^2}{\omega _1^2-\omega _2^2}+\frac{\left(1-r^2\right) \left(r \pi _r \cos\xi -\pi _{\xi } \sin\xi\right){}^2}{r^2 \left(\omega _1^2-\omega _3^2\right)},\\
F_{2}&=r^2 \sin ^2\xi+\frac{\pi _{\xi }^2}{\omega _2^2-\omega _1^2}+\frac{\left(1-r^2\right) \left(\pi _{\xi } \cos\xi+r \pi _r \sin\xi\right){}^2}{r^2 \left(\omega _2^2-\omega _3^2\right)},\\
F_{3}&=1-r^{2}+\frac{\left(1-r^2\right) \left(r \pi _r \cos\xi-\pi _{\xi } \sin\xi\right){}^2}{r^2 \left(\omega _3^2-\omega _1^2\right)}+\frac{\left(1-r^2\right) \left(\pi _{\xi } \cos\xi+r \pi _r \sin\xi\right){}^2}{r^2 \left(\omega _3^2-\omega _2^2\right)}.
\end{align}
In order to write the deformed conserved quantities in a compact form, we will first define the following functions:
\begin{align*}
s_{1}=&\pi _\xi ^2\left( {1 - \frac{1}{{{r^2}}}} \right){\omega _1}\left( {2r{\pi _r}\cos 2\xi  - {\pi _\xi }\sin 2\xi } \right) - {\pi _\xi }\pi _r^2\left( {1 - {r^2}} \right){\omega _1}\sin 2\xi,\\ 
s_{2}=&- {\pi _\xi }{\pi _r}r\left( {1 - {r^2}} \right)\left( {\omega _1^2 + 2{\omega _3}{\omega _1} - \omega _2^2} \right)\sin 2\xi  - \pi _\xi ^2\pi _r^2\left( {1 - {r^2}} \right){\cos ^2}\xi\\ &- \pi _\xi ^3\left( {1 - \frac{1}{{{r^2}}}} \right)\left( {2r{\pi _r}\cos \xi  - {\pi _\xi }\sin \xi } \right)\sin \xi\\
& - \pi _\xi ^2\left[ {\omega _3^2\left( {1 - {r^2}{{\cos }^2}\xi } \right) + \left( {1 - {r^2}} \right)\left( {\omega _1^2{{\cos }^2}\xi  + \omega _2^2{{\sin }^2}\xi  + 2{\omega _1}{\omega _3}\cos 2\xi } \right)} \right],\\
s_{3}=&- 2{\pi _\xi }\left( {1 - {r^2}} \right)\left( {r{\pi _r}\cos \xi  - {\pi _\xi }\sin \xi } \right)\left[ {{\pi _\xi }\left( {{\omega _1} + {\omega _3}} \right)\cos \xi  + r{\pi _r}{\omega _1}\sin \xi } \right],\\ 
s_{4}=& - \pi _\xi ^2\left( {1 - {r^2}} \right){\left( {r{\pi _r}\cos \xi  - {\pi _\xi }\sin \xi } \right)^2}.
\end{align*}
Using these functions, the conserved quantities for the deformed model can be written in these coordinates as:
\begin{align}
 {\mathscr F}_{1}=&F_{1}+\frac{\sum\limits_{i = 1}^4 {{s_i}{\varkappa ^i}} }{{\left( {\omega _1^2 - \omega _2^2} \right)\left( {\omega _1^2 - \omega _3^2} \right)}}+\frac{{{\varkappa ^2}\pi _\xi ^2{r^2}\omega _1^2{{\sin }^2}\xi }}{{\left( {\omega _1^2 - \omega _2^2} \right)\left( {\omega _1^2 - \omega _3^2} \right)}} + \frac{{{\varkappa ^2}\pi _r^2{r^2}\left( {1 - {r^2}} \right){{\cos }^2}\xi }}{{\omega _1^2 - \omega _3^2}}\\ 
&+ \frac{{2\varkappa {\pi _r}r\left( {1 - {r^2}} \right){\omega _3}{{\cos }^2}\xi }}{{\omega _1^2 - \omega _3^2}} - \varkappa {\pi _\xi }\left( {\frac{{{r^2}{\omega _1}}}{{\omega _1^2 - \omega _2^2}} + \frac{{\left( {1 - {r^2}} \right){\omega _3}}}{{\omega _1^2 - \omega _3^2}}} \right)\sin 2\xi \nonumber,
\end{align}
\begin{align}
 {\mathscr F}_{2}=&F_{2}  + \frac{{\sum\limits_{i = 1}^4 {{s_i}{\varkappa^i}} }}{{\left( {\omega _2^2 - \omega _1^2} \right)\left( {\omega _2^2 - \omega _3^2} \right)}} + \frac{{{\varkappa ^2}\pi _\xi ^2{r^2}\omega _2^2{{\sin }^2}\xi }}{{\left( {\omega _2^2 - \omega _1^2} \right)\left( {\omega _2^2 - \omega _3^2} \right)}} + \frac{{{\varkappa ^2}\pi _r^2{r^2}\left( {1 - {r^2}} \right){{\sin }^2}\xi }}{{\omega _2^2 - \omega _3^2}}\\
&+ \frac{{2\varkappa {\pi _r}r\left( {1 - {r^2}} \right){\omega _3}{{\sin }^2}\xi }}{{\omega _2^2 - \omega _3^2}}\ - \varkappa {\pi _\xi }\left( {\frac{{{r^2}{\omega _1}}}{{\omega _2^2 - \omega _1^2}} - \frac{{\left( {1 - {r^2}} \right){\omega _3}}}{{\omega _2^2 - \omega _3^2}}} \right)\sin 2\xi,\nonumber\\
 {\mathscr F}_{3}=&F_{3} + \frac{{\sum\limits_{i = 1}^4 {{s_i}{\varkappa ^i}} }}{{\left( {\omega _3^2 - \omega _1^2} \right)\left( {\omega _3^2 - \omega _2^2} \right)}} + \frac{{{\varkappa ^2}\pi _\xi ^2{r^2}\omega _3^2{{\sin }^2}\xi }}{{\left( {\omega _3^2 - \omega _1^2} \right)\left( {\omega _3^2 - \omega _2^2} \right)}}\\
&+ {\varkappa ^2}\pi _r^2{r^2}\left( {1 - {r^2}} \right)\left( {\frac{{{{\cos }^2}\xi }}{{\omega _3^2 - \omega _1^2}} + \frac{{{{\sin }^2}\xi }}{{\omega _3^2 - \omega _2^2}}} \right){ - \frac{{\varkappa {\pi _\xi }\left( {1 - {r^2}} \right)\left( {\omega _1^2 - \omega _2^2} \right){\omega _3}\sin 2\xi }}{{\left( {\omega _3^2 - \omega _1^2} \right)\left( {\omega _3^2 - \omega _2^2} \right)}}}\nonumber\\
&+ 2\varkappa \pi_{r} r\left( {1 - {r^2}} \right){\omega _3}\left( {\frac{{{{\cos }^2}\xi }}{{\omega _3^2 - \omega _1^2}} + \frac{{{{\sin }^2}\xi }}{{\omega _3^2 - \omega _2^2}}} \right)\nonumber.
\end{align}


\end{document}